# First Application of Large Reactivity Measurement through Rod Drop Based on Three-Dimensional Space-Time Dynamics

Wencong Wang, Liyuan Huang, Caixue Liu, Han Feng, Jiang Niu, Qidong Dai,

Linfeng Yang, Mingchang Wu

Nuclear Power Institute of China, China

**Abstract**

Reactivity measurement is an essential part of a zero-power physics test, which is critical to reactor design and development. The rod drop experimental technique is used to measure the control rod worth in a zero-power physics test. The conventional rod drop experimental technique is limited by the spatial effect and the difference between the calculated static reactivity and measured dynamic reactivity; thus, the method must be improved. In this study, a modified rod drop experimental technique that constrains the detector neutron flux shape function based on three-dimensional space-time dynamics to reduce the reactivity perturbation and a new method for calculating the detector neutron flux shape function are proposed. Correction factors were determined using Monte Carlo N-Particle transport code and transient analysis code for a pressurized water reactor at the Ulsan National Institute of Science and Technology and Xi'an Jiaotong University, and a large reactivity of over 2000 pcm was measured using the modified technique. This research evaluated the modified technique accuracy, studied the influence of the correction factors on the modification, and investigated the effect of constraining the shape function on the reactivity perturbation reduction caused by the difference between the calculated neutron flux and true value, using the new method to calculate the shape function of the detector neutron flux and avoiding the neutron detector response function (weighting factor) calculation.

# 1. Introduction

Reactivity measurement is an essential part of a zero-power physics test, which is vital for reactor design and development. In a zero-power physics test, different core configuration types are investigated, and the reactor power is limited to a low level; thus, the ex-core detector signal level is also low. Therefore, it is critical to determine the reactivity with a low-strength signal, especially when the reactivity is large.

The rod swap and boron dilution methods can be used to measure the reactivity in a zero-power physics test. The rod swap method is slow, and it is easily influenced by the control rod shadow effect. The boron dilution method is accurate; however, it takes a long time to dilute the boron to compensate for the reactivity loss. This method is typically applied in commercial nuclear power plants, but it is rarely applied in a zero-power critical reactor because it is difficult to dilute boron in an experimental reactor.

In recent decades, dynamic reactivity measurement methods have been widely used in commercial nuclear power plant startups. They separately and independently measure the worth of each control rod bank by inserting each control rod bank into the core at its maximum allowable speed.

Dynamic methods, such as the dynamic rod worth measurement(DRWM) method developed by Chao, [1–3] dynamic reactivity measurement of rod worth method developed by Kastanya et al., [4] and the dynamic control rod reactivity measurement method developed by Lee et al., [5,6] have exhibited excellent results for numerous pressurized water reactor (PWR) startups. In these studies, the three-dimensional (3D) space-time kinetics theory is employed to overcome the limitations of a one-point reactor model in dynamic measurement. The reactivities measured by inserting and withdrawing the control rod bank at its maximum allowable speed in a commercial nuclear power plant were all less than 2000 pcm. Recently, Sang Ji Kim et al. [7] presented a preliminary dynamic rod drop simulation study for a transuranic burner core mockup of a sodium-cooled fast reactor, in which the reactivities were approximately 1000 pcm. However, there were several difficulties in applying these methods in a zero-power experimental reactor.

(1) The ex-core detector signal level was lower than that of a commercial nuclear power plant due to the small core configurations. Hence, with the control rod inserted at its maximum allowable speed, the detector signal can easily fall below the lower limit of the measurement.

(2) In certain cases, the reactivity of a single control rod is much larger than that of a commercial nuclear power plant control rod bank. For such a large reactivity insertion, the neutron flux distributions in the core and ex-core detector change rapidly and substantially, which may increase the difference between the calculated neutron flux and its true value, affecting the accuracy of the correction factors and reactivity results.

Due to these difficulties, the DRWM method [1–9] is not always the optimal choice for the zero-power physics test of a zero-power experimental reactor. Thus, in certain cases, the rod drop experimental technique is used to measure the reactivity in a zero-power physics test. [10]

The rod drop experimental technique measures the reactivity by dropping the control rod into the core, and the reactivity is analyzed using a one-point reactor model, which assumes that the neutron flux distribution in the reactor core maintains the same shape during the dynamic measurement. However, for large reactivity insertions, the neutron flux distribution changes greatly during the dynamic process, which can significantly affect the detector signal. Thus, the accuracy of a large reactivity measured during this process is unsatisfactory, indicating that the rod drop experimental technique must be modified.

In this study, a modified rod drop experimental technique was applied to a large reactivity insertion measurement to reduce the measurement error. We propose a method that constrains the shape function of the detector neutron flux based on 3D space-time dynamics [11] to reduce the reactivity perturbation caused by calculated correction factor errors, which is affected by the difference between the calculated neutron flux and its true value. We also propose a new method to calculate the shape function of the detector neutron flux that avoids the neutron detector response function (weighting factor) calculation. Finally, we evaluate the proposed method accuracy, influence of the correction factors on the modification, and effect of constraining the shape function of the detector neutron flux.

Compared to the previous research, our work is innovative in that we are the first to apply the modification of the rod drop experimental technique to a large reactivity measurement of over 2000 pcm in a zero-power experimental reactor, to constrain the shape function to reduce the reactivity perturbation caused by the calculated correction factor errors, and to propose a new method to calculate the detector neutron flux shape function without calculating the neutron detector response function (weighting factor).

The difficulties of the conventional rod drop experimental technique and the modified method theory are discussed in Section 2. The proposed method application to large reactivity measurement is introduced in Section 3. The results and are described and discussed in Section 4. The conclusions are summarized in Section 5.

**2. Methodology**

2.1 Conventional rod drop experimental technique

The usual rod-drop method was analyzed according to the prompt jump theory. Its practical difficulty lies in determining the prompt jump flux level because the actual reactivity insertion is gradually terminated, not in a step. Therefore, the prompt flux decrease during the latter phase of the reactivity insertion overlaps with the beginning of the flux decrease due to delayed neutron source decay. Improved accuracy requires special correction techniques or an analysis of the inverse kinetics. [11]

Thus, in this research, the inverse kinetics based on a one-point reactor model was employed to analyze the data measured by dropping a control rod into a core, which we defined as the conventional rod drop experimental technique to make more easily distinguish the method before and after modification.

The measurement procedure is as follows:

(1) Operate the reactor core in the critical state.

(2) Drop the control rod into the core while recording the neutron signal with the ex-core neutron detector.

(3) Process the neutron signal using reactivity measurement equipment.

(4) Calculate the reactivity using the experiment data and inverse kinetic equation.

The inverse kinetic equation can be described as follows: [12]

$$\rho = \beta + \Lambda \frac{dN(t)}{N(t)dt} - \frac{1}{N(t)}\sum_{i=1}^{6}\beta_i \lambda_i \int_{t_0}^{t} N(\tau) e^{-\lambda_i(t-\tau)} d\tau , \qquad (1)$$

where $\rho$ is the reactivity, $\beta$ is the effective fraction of delayed neutron, $\Lambda$ is the neutron generation time, and $\lambda$ is the delayed neutron decay constant. $\beta$, $\Lambda$, and $\lambda$ are calculated before the experiment, and they are considered to be known constant parameters in the measurement. The ENDF/B7.0 cross-section libraries were used to calculate $\beta$ and $\lambda$ in this research. $N(t)$ is the neutron detector signal, which was assumed to be proportional to the amplitude function $p(t)$; therefore, the amplitude function $p(t)$ was replaced with $N(t)$ in Equation (1). There are certain difficulties in using the conventional method, such as the spatial effect of the neutron detector signal and difference between static reactivity and dynamic reactivity.

(1) Spatial effect of the neutron detector signal

In a one-point reactor model, the time dependence of the flux $\phi(r,E,t)$ is assumed to be separable from its space and energy dependences, and the flux can be described as the product of amplitude function $p(t)$ and shape function $\psi(r,E)$, assuming that the shape function is unchanged during measurement:

$$\phi(r,E,t) = p(t)\psi(r,E) . \qquad (2)$$

With this assumption, the shape function $\psi(r,E)$ is independent of time, and the neutron flux $\phi(r,E,t)$ is proportional to the amplitude function $p(t)$ in the measurement; therefore, the detector neutron flux $\phi(r_d,E,t)$ is proportional to the amplitude function $p(t)$.

In a real situation, the shape function of the flux is time-dependent, and it changes significantly during the rod drop process. Thus, the detector neutron flux is affected by changes in the neutron flux shape. This violates the assumption that the shape function is unchanged during measurement. For this reason, the detected neutron signal is not proportional to the neutron flux amplitude function, and using it to determine the reactivity leads to measurement inaccuracies.

(2) Difference between static reactivity and dynamic reactivity

The reactivity determined by the inverse kinetic equation is called the dynamic reactivity. It is definitional different from the reactivity determined by static calculation, which is called the static reactivity. Thus, to obtain superior results, it is important to remedy these difficulties in rod drop reactivity measurement.

2.2 Modified rod drop experimental technique

The modified rod drop experimental technique is based on exact point dynamics, which are the 3D reactor dynamics. The modification is focused on two aspects: (1) the difference between the neutron flux amplitude function and neutron detector signal and (2) the difference between the calculated static reactivity and measured dynamic reactivity.

2.2.1 Detector signal correction

2.2.1.1 Detector signal correction factor

From the exact point dynamics, the neutron flux in the reactor core $\phi(r,E,t)$ can be factorized into a purely time-dependent amplitude function, $p(t)$, and a space-, energy-, and time-dependent neutron flux shape function $\psi(r,E,t)$ : [11]

$$\phi(r,E,t) = p(t)\psi(r,E,t) \ . \tag{3}$$

Based on the exact point dynamics, the neutron signal $N_{Det}(t)$ is related to the neutron flux in the reactor core $\phi(r,E,t)$ : [11]

$$N_{Det}(t) = \iint_{V\ E} W(r,E)\phi(r,E,t)dEdV \ , \tag{4}$$

where $W(r,E)$ is the weighting factor that denotes the neutron contribution degree at position $r$ in the core region to the neutron detector signal, describing the space and energy dependences of the neutron detector sensitivity.

Thus, Equation (5) can be written as follows:

$$N_{Det}(t) = \int_V \int_E W(r,E)p(t)\psi(r,E,t)dEdV \quad . \tag{5}$$

The neutron flux amplitude function $p(t)$ can be extracted from the neutron signal using the following equation:

$$p(t) = \frac{N_{Det}(t)}{\int_V \int_E W(r,E)\psi(r,E,t)dEdV} \quad . \tag{6}$$

Using Equation (6), the neutron signal is converted into the neutron flux amplitude function $p(t)$, and the difference between the neutron flux shape and neutron signal can be largely reduced, resulting in a more accurate result.

The neutron signal $N_{Det}(t)$ can also be described as follows:

$$N_{Det}(t) = \Sigma(t)\phi(r_d,t) = \Sigma(t)p(t)\psi(r_d,t) \quad , \tag{7}$$

where $\Sigma(t)$ is the detector sensitivity.

From Equations (6) and (7):

$$\int_V \int_E W(r,E)\psi(r,E,t)dEdV = \Sigma(t)\psi(r_d,t) \quad . \tag{8}$$

Thus, the denominator on the right side of Equation (6) is considered to be proportional to the shape part of the detector neutron flux $\psi(r_d,t)$.

To modify the detector signal, one must calculate the denominator on the right side of Equation (6). The neutron detector response function (weighting factor) $W(r,E)$ can reduce the accuracy due to the mesh precision and the assumption of no changes under different control rod patterns, and the denominator on the right side of Equation (6) is proportional to the shape part of the detector neutron flux $\psi(r_d,t)$. Thus, we use the shape part of the detector neutron flux $\psi(r_d,t)$ to describe the denominator on the right side of Equation (6) to reduce the inaccuracy induced by the neutron detector response function (weighting factor) $W(r,E)$. The new method for calculating the shape part of the detector neutron flux $\psi(r_d,t)$ is described in detail in Section 3.2.5.

Normalization about the detector signal is performed as follows:

$$\overline{p(t)} = \frac{p(t)}{p(0)} = \frac{N_{Det}(t)}{N_{Det}(0)} \cdot \frac{\Sigma(0)}{\Sigma(t)} \cdot \frac{1}{C_{det}} = \frac{N_{Det}(t)}{N_{Det}(0)} \cdot \frac{1}{C_{det}} \quad . \tag{9}$$

Because the calculated energy spectra of the detector neutron flux stay the same during the rod drop process, $\Sigma(t)$ is assumed to be unchanged in the measurement.

$$C_{det} = \frac{\psi(r_d,t)}{\psi(r_d,0)} \quad , \tag{10}$$

where $C_{det}$ is the detector signal correction factor, and "0" refers to the critical state at the beginning of the rod drop process. By substituting the normalized amplitude function $\overline{p(t)}$ into Equation (1), instead of the neutron detector signal $N(t)$, the reactivity $\rho$ is obtained.

2.2.1.2 Constraining the shape function

To modify the detector signal, the shape part of the detector neutron flux $\psi(r_d,t)$ should be calculated. The purpose of constraining the shape function is to reduce the reactivity perturbation caused by the neutron flux or shape function calculation error.

We begin with the definition of $\psi(r,E,t)$, which is the neutron flux shape function.

From the 3D neutron diffusion equation:

$$\frac{1}{v}\frac{\partial \phi(r,E,t)}{\partial t} = (F_p - M)\phi(r,E,t) + S_d(r,E,t) + S(r,E,t) \quad , \tag{11}$$

where $M$ is the neutron destruction operator, $F_p$ is the prompt neutron production operator, $S_d(r,E,t)$ is the delayed neutron source, and $S(r,E,t)$ is an independent source.

By substituting Equation (3) into Equation (11), we obtain the following

equation:

$$\frac{1}{v}[\dot{p}(t)\psi(r,E,t) + p(t)\dot{\psi}(r,E,t)] = (F_p - M)p(t)\psi(r,E,t) + S_d(r,E,t) + S(r,E,t)$$

. (12)

Allowing the shape function to depend on time is a first generalization compared to the use of the time-independent shape in the derivation presented in Equation (2). A second generalization used in the derivation does not remove an approximation, but rather exploits a certain freedom of choice; the neutronics equation is multiplied by a weight function, $\phi^w(r,E)$, prior to integration with respect to space and energy. The flux factorization is also introduced into the left side of equation (11). [11]

$$\frac{dp(t)}{dt}\int\int_{V\ E}\frac{\phi^w(r,E)\psi(r,E,t)}{v(E)}dEdV + p(t)\frac{d}{dt}\int\int_{V\ E}\frac{\phi^w(r,E)\psi(r,E,t)}{v(E)}dEdV$$
$$= \int\int_{V\ E}\phi^w(F_p - M)p(t)\psi(r,E,t)dEdV + \int\int_{V\ E}\phi^w S_d(r,E,t)dEdV + \int\int_{V\ E}\phi^w S(r,E,t)dEdV$$

. (13)

The second term on the left side of Equation (13) that appears with flux factorization can be eliminated by only using the integral to constrain the time variation of the shape function, thus making the factorization unique:

$$\int_V\int_0^\infty \frac{\phi^w(r,E)\psi(r,E,t)}{v(E)}dEdV = C \quad, \tag{14}$$

where $C$ is an arbitrary constant.

Constraining the neutron flux shape using Equation (14) does not introduce an approximation, and the factorization introduces a new degree of freedom. With the constraint defined as Equation (14), Equation (13) performs the same as the one-point model without any assumptions or approximations. Thus, the constraint is vital in determining the neutron flux shape function $\psi(r,E,t)$.

The $\phi^w(r,E)$ choice is also important because it can affect the $\psi(r,E,t)$ accuracy and thus affect the detector signal modification.

In practice, the shape function $\psi(r,E,t)$ cannot be precisely determined (the calculation result is always somewhat different from the true value), and the calculated flux and flux amplitude $p(t)$ are therefore only approximate solutions. Both values depend on the weight function. It is thus advantageous to choose a weight function that reduces the error resulting from inaccuracies in the shape function. Because the solution of the point kinetics equation is particularly sensitive to reactivity errors, a weight function should be selected that reduces the effect of shape function inaccuracies on the reactivity.

The initial adjoint flux, $\phi_0^*(r,E)$, fulfills this objective. [11] This is because, in static perturbation theory, the first order dominates the reactivity perturbation, which is caused by the perturbation in neutron flux, and therefore, the use of $\phi_0^*(r,E)$ can eliminate the first order in the reactivity perturbation. Thus, the use of $\phi_0^*(r,E)$ can eliminate most of the reactivity perturbation caused by the difference between the calculated neutron flux and true value, reducing the reactivity perturbation caused by the calculation error of the neutron flux or shape function.

It is more precise to use $\phi^*(r,E,t)$ as the weight function $\phi^w(r,E)$. However, $\phi^w(r,E)$ should remain unchanged in the measurement to maintain the factorization consistency (Equation (3)). Thus, the use of $\phi_0^*(r,E)$ is the optimal choice because it can eliminate the reactivity perturbation caused by the error in the neutron flux calculation.

In summary, to modify the detector signal, it is essential to impose a constraint on the shape function, making the factorization unique with the initial adjoint flux $\phi_0^*(r,E)$ as the weighting function.

$$\int_V \int_0^\infty \frac{\phi_0^*(r,E)\psi(r,E,t)}{v(E)} dEdV = K_0 \ , \tag{15}$$

where $\phi_0^*(r,E)$ is the adjoint neutron flux distribution of the critical state at the beginning of the rod drop measurement, $v(E)$ is the neutron velocity, and $K_0$ is an arbitrary constant, which is 1 in this study.

2.2.3 Dynamic reactivity correction

The reactivity determined by the inverse kinetic Equation (1) is typically considered the dynamic reactivity, which differs from the static reactivity. To remedy the difference between calculated static reactivity and measured dynamic reactivity, a dynamic reactivity correction is performed as follows:

$$\rho_{st,m} = C_{dyn} \cdot \rho_{dyn,m} \ , \tag{16}$$

$$C_{dyn} = \frac{\rho_{st,c}}{\rho_{dyn,c}} ,  \qquad (17)$$

where $\rho_{dyn,m}$ is the measured dynamic reactivity, $\rho_{st,c}$ is the calculated static reactivity, and $\rho_{dyn,c}$ is the calculated dynamic reactivity, which is determined by substituting the calculated neutron flux amplitude function into the inverse kinetic equation.

To perform the correction, the difference between the calculated static reactivity and measured dynamic reactivity is determined via theoretical calculation. The static reactivity $\rho_{st,c}$ is determined using a static eigenvalue calculation, and the neutron flux amplitude function is simulated using a transient calculation code. By substituting the simulated neutron flux amplitude function into Equation (1), the calculated dynamic reactivity $\rho_{dyn,c}$ is obtained. Then, the dynamic reactivity correction factor is determined by incorporating the calculated static reactivity and calculated dynamic reactivity into Equation (17). In this research, the static reactivity is calculated using Monte Carlo N-Particle (MCNP) transport code [13], and the neutron flux amplitude function is simulated transient analysis code for a pressurized water reactor at the Ulsan National Institute of Science and Technology and Xi'an Jiaotong University (TAPUX) [14].

## 3. Application

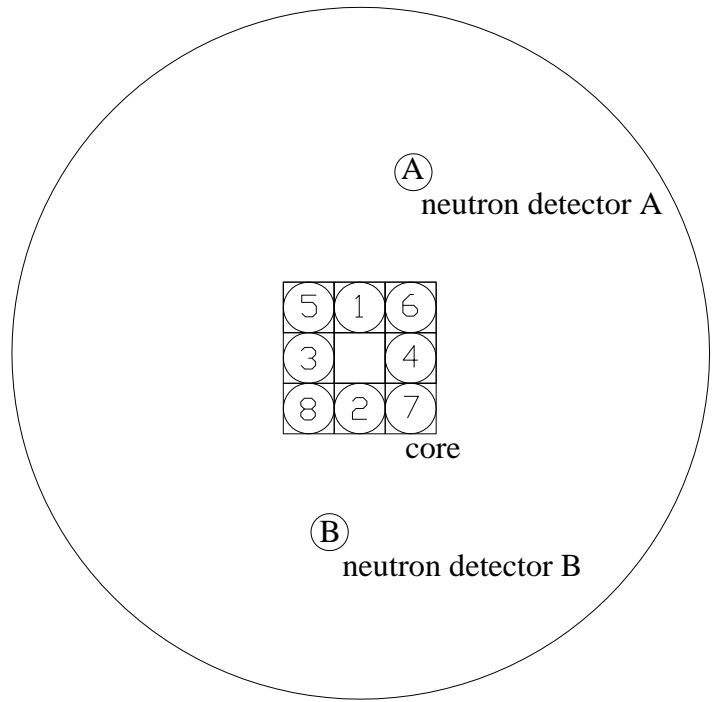

Fig. 1. Schematic of experimental core configuration.

In this study, the reactivity measurement of a zero-power water-moderated thermal critical reactor was performed. Strong reactivity was locally added into the core by dropping a single control rod cluster into the core from its full out state when the reactor operates in a critical state.

Two control rod clusters (#1 and #7) were chosen to drop from the critical state in this research, and the experimental core configuration is illustrated in Fig. 1. Two gamma-compensated neutron ionization chambers, illustrated by circles A and B in Fig. 1, were used to measure the neutron signal, which was located outside the reactor core,. Neutron detector A was used to measure the neutron signal of control rod cluster #1 dropping from the critical control rod pattern 1, and neutron detector B was used to measure control rod cluster #7 dropping from critical control rod pattern 2 (Table 1). The detector configuration was based on experimental experience, and the detector correction factors of detector A were large, while those of detector B were small from the modification perspective.

A flow diagram of the modified rod drop experimental technique is displayed in Fig. 2. To modify the measurement, both static and dynamic calculations should be performed according to the following steps:

(1) Build 3D MCNP and 3D dynamic calculation models.

(2) Calculate the static physical parameters.

(3) Constrain the detector neutron flux into the shape function, and modify the measured neutron signal.

(4) Calculate the amplitude function $p(t)$ of the dynamic rod drop process.

(5) Obtain the dynamic reactivity correction factor to determine the measured final reactivity.

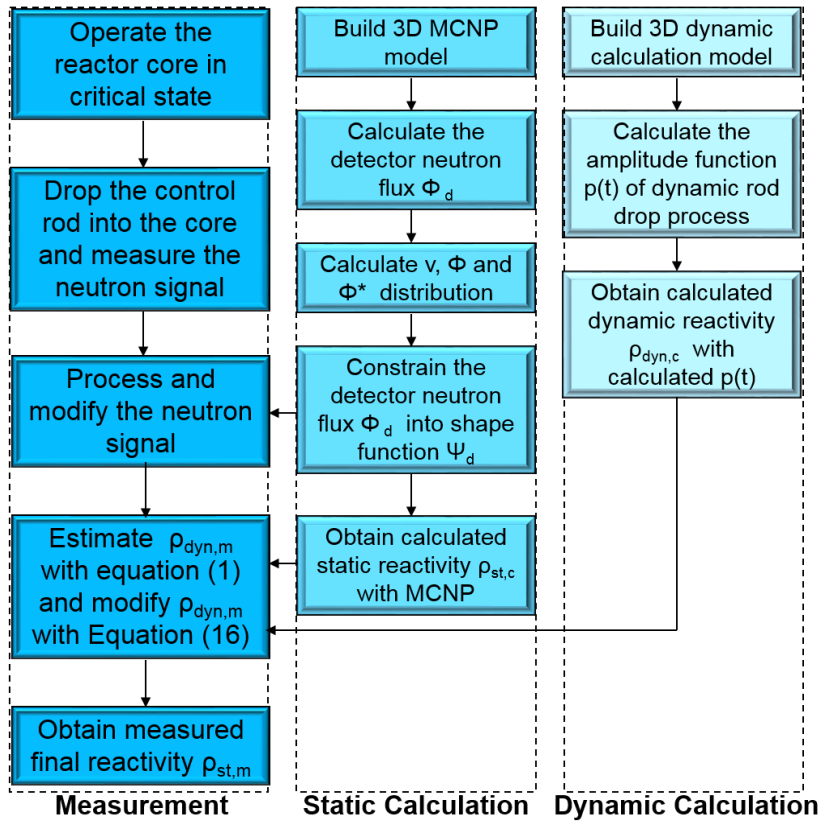

Fig. 2. Flow diagram of modified rod drop experimental technique.

3.1 Calculation model

Table 1. Results of experimental critical control rod pattern calculations using MCNP

| Number | Experimental critical control rod pattern | MCNP keff ± error |
|---|---|---|
| 1 | #1–4 rods on top and #5–8 rods at same critical position | 0.99975±0.00005 |
| 2 | #1, #2, and #5–8 rods on top, and #3 and #4 rods at same critical position | 0.99980±0.00005 |
| 3 | #1–5 rods on top, #7 rod at critical position, and #6 and #8 rods at bottom | 0.99971±0.00005 |

The calculation geometry is based on Fig. 1. Both the MCNP and dynamic calculation models used for modification in this research were verified by experimental critical control rod pattern. The results of the experimental critical control rod pattern calculation by MCNP are presented in Table 1. The ENDF/B-VI cross-section libraries were used in the MCNP code.

The detector was located outside the core, and the calculation result of the detector neutron flux energy spectra remained the same before and after the rod drop. Therefore, the neutron detector efficiency $\Sigma$ was assumed to be constant in the rod drop process. Because the shape function at detector position $\psi(r_d,t)$ appears a as ratio in Equation (10), the neutron detector in the model was simplified as a cylinder with its shell, and each detector was located in a tube. The detector was located approximately 10 cm above the bottom of the active core in the axial direction. The sensitive height and radius of the detector were approximately 36 cm and 2.5 cm, respectively.

In this study, all of the static physical parameters were determined using the MCNP code, such as the detector neutron flux $\phi(r_d,t)$, neutron velocity $v$, neutron flux $\phi(r,E,t)$, and adjoint neutron flux distribution in the core $\phi_0^*(r,E)$. The dynamic physical parameter, which is the amplitude function $p(t)$ in the rod drop process, was calculated using TAPUX.

3.2 Calculation of static physical parameters

3.2.1 Detector neutron flux

The detector neutron flux $\phi(r_d,t)$ was calculated using an MCNP neutron flux tally card based on the previously mentioned MCNP calculation model. The geometry split technique was used to improve the calculation efficiency. The calculated detector neutron flux $\phi(r_d,t)$ was constrained into a shape function at detector position $\psi(r_d,t)$, which is elaborated on in Section 3.2.5.

3.2.2 Neutron flux distribution

The neutron flux distribution calculation in the core $\phi(r,E,t)$ was directly performed using the neutron flux tally card. The volume tally mesh was in the XY plane assembly-wise and 10 cm in the Z-direction.

3.2.3 Neutron velocity

The calculation of neutron velocity $v$ in the core was realized using the tally energy card and MCNP neutron flux tally card. The volume tally mesh was the same as that of the neutron flux distribution calculation. Using the calculated $\phi(r,E,t)$ of each energy bin in each volume mesh and the relationship between energy and velocity, the neutron velocity $v$ was calculated as follows:

$$\bar{E}(t) = \frac{\sum_{j=1}^{n_V}\sum_{i=1}^{n_E}\frac{E_{i,up}+E_{i,down}}{2}\phi_{i,j}}{\sum_{j=1}^{n_V}\sum_{i=1}^{n_E}\phi_{i,j}}, \quad (18)$$

$$v(t) = \sqrt{\frac{2\bar{E}(t)}{m}}. \quad (19)$$

The average neutron energy in core $\bar{E}$ was determined by the upper and lower energy boundaries and the corresponding neutron flux of each energy bin in the volume tally mesh elements. $n_E$ is the number of energy bins in a volume tally mesh element, and it was the same for all of the $n_V$ volume tally mesh elements.

3.2.4 Adjoint neutron flux

To constrain the detector neutron flux $\phi(r_d,t)$ into a shape function at the detector position $\psi(r_d,t)$, the adjoint neutron flux distribution in the core $\phi_0^*(r,E)$ was calculated.

The proportionality of the iterated fission probability ($I_{FP}$) to the adjoint flux was demonstrated [15], and $I_{FP}$ was therefore calculated instead of the adjoint neutron flux $\phi_0^*(r,E)$, which was difficult to obtain using the MCNP code. $I_{FP}$ was determined using the formulas below [15,16]:

$$I_{FP}^{(\lambda')}(\theta_0) = \frac{s_{\theta_0}^{(1)}}{s_{\theta_0}^{(0)}}\frac{s_{\theta_0}^{(2)}}{s_{\theta_0}^{(1)}}\cdots\frac{s_{\theta_0}^{(\lambda'-1)}}{s_{\theta_0}^{(\lambda'-2)}}\frac{s_{\theta_0}^{(\lambda')}}{s_{\theta_0}^{(\lambda'-1)}}, \quad (20)$$

$$I_{FP}^{(\lambda')}(\theta_0) = k_{\theta_0}^{(1)}k_{\theta_0}^{(2)}\cdots k_{\theta_0}^{(\lambda'-1)}k_{\theta_0}^{(\lambda')}. \quad (21)$$

As the generation number, $\lambda'$, increases, distributions of the neutron flux $\phi^{(\lambda)}$ and the fission neutron emission (source) produced by the progenies converge to those of the fundamental mode. $s_{\theta_0}^{(i)}$ is the number of fission neutrons of the i-th generation; hence, the corresponding ratio $s_{\theta_0}^{(i)} / s_{\theta_0}^{(i-1)}$ can be written as $k_{\theta_0}^{(i)}$, which can be estimated by the eigenvalue calculation in MCNP.[15] $\theta_0$ is the initial point source of $I_{FP}$, and the $I_{FP}$ of $\theta_0 = (r_0, E_0, \Omega_0)$ can be estimated by calculating the initial source points at $\theta_0$. The $I_{FP}$ was calculated using Equation (21).

In the $I_{FP}$ calculation, $k_{\theta_0}^{(i)}$ was determined using k-collision, which was printed in the MCNP output file. The $I_{FP}$ of every tally mesh element was determined using a corresponding eigenvalue calculation. In every eigenvalue calculation, sufficient neutrons were sampled by locating hundreds of homogenous source points in the corresponding mesh grid.

3.2.5 Shape function at detector position $\psi(r_d, t)$

The new method for calculating the shape part of the detector neutron flux $\psi(r_d, t)$ is described in detail in this section. $\psi(r_d, t)$ was determined using the detector neutron flux $\phi(r_d, t)$ and constraint factor of the shape function $C(t)$.

For the static physical parameters calculated using the methods described in Sections 3.2.2–3.2.4, the neutron flux in the core $\phi(r, E, t)$ was constrained into shape function in the core $\psi(r, E, t)$ as follows:

$$\psi(r, E, t) = \frac{\phi(r, E, t)}{C(t)}, \tag{22}$$

where $C(t)$ is the constraint factor of the shape function:

$$C(t) = \int_V \int_0^\infty \frac{\phi_0^*(r, E) \phi(r, E, t)}{v(E)} dE dV. \tag{23}$$

The shape function satisfies the constraint condition Equation (15) when $K_0 = 1$.

The detector neutron flux $\phi(r_d, t)$ was constrained into shape function at detector position $\psi(r_d,t)$ as follows.

From Equation (8):

$$\psi(r_d,t) = \int_V \int_E W(r,E)\psi(r,E,t)dEdV / \Sigma(t) \quad . \tag{24}$$

By substituting Equation (22) into Equation (24):

$$\psi(r_d,t) = \frac{\int_V \int_E W(r,E)\phi(r,E,t)dEdV / \Sigma(t)}{C(t)} \quad . \tag{25}$$

Thus, $\psi(r_d,t)$ can be described as

$$\psi(r_d,t) = \frac{\phi(r_d,t)}{C(t)} \quad . \tag{26}$$

$$\psi(r_d,t) = \frac{\phi(r_d,t)}{\int_V \int_0^\infty \frac{\phi_0^*(r,E)\phi(r,E,t)}{v(E)} dEdV} \quad . \tag{27}$$

As displayed in Equation (27), the shape function at detector position $\psi(r_d,t)$ can be determined using the detector neutron flux $\phi(r_d,t)$ and constraint factor of the shape function $C(t)$. Hence, the weighting factor $W(r,E)$ calculation is avoided.

This method appears to be superior to that calculating the neutron detector response function (weighting factor) $W(r,E)$ [17,18] because $W(r,E)$ can reduce the accuracy due to the mesh precision and assumption that it does not change under different control rod patterns.

3.3 Dynamic physical parameter calculations

The amplitude function $p(t)$ was determined using TAPUX and the dynamic model mentioned above.

The TAPUX code is a recently developed 3D two-group light water reactor core analysis code by the Nuclear Engineering Computational Physics laboratory at Xi'an Jiaotong University. It can be used by utilities to perform transient analysis in neutronics. The code adopts the non-linear coarse-mesh finite difference method based on the nodal methodology in steady-state and transient core calculations. The frequency transform method was applied based on the θ method in time discretization. Thus, the time step can be expanded to enhance the efficiency.

The cross-section of TAPUX was derived from the Bamboo lattice code [19–21], which was based on the ENDF/B7.0 library and 69-group lattice calculation. A two-group homogenized cross-section was generated for each assembly and made into a look-up table for the transient calculation in TAPUX considering the fuel temperature feedback, coolant density, and control rod movements.

For the calculation, the control rod was dropped into the core for free fall, which was the same as that in a real experiment situation. Thus, the amplitude function $p(t)$ in the rod drop process was determined.

3.4 Correction factor calculations

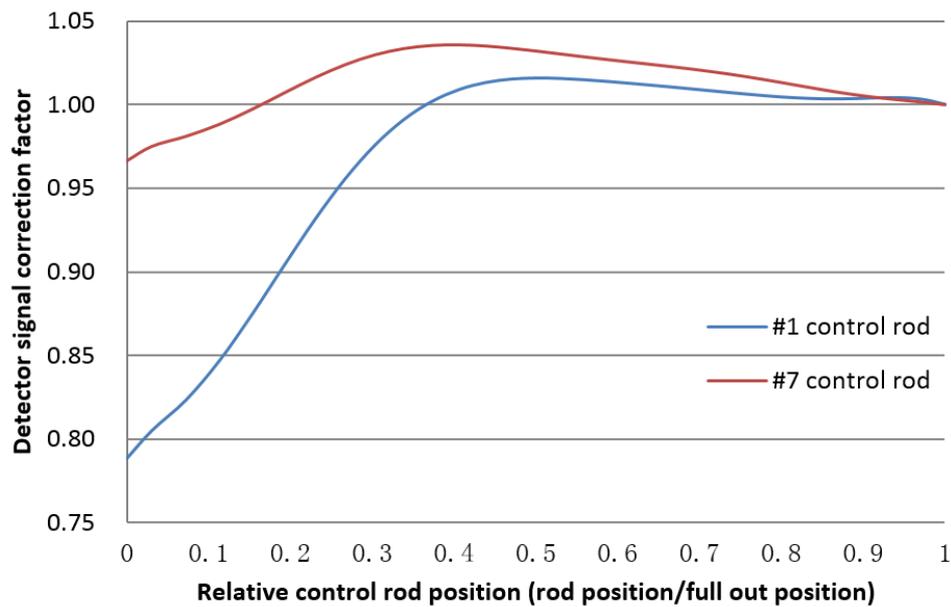

Fig. 3. Detector signal correction factor.

The shape function at detector position $\psi(r_d, t)$ was determined using the static physical parameters, which were calculated using the MCNP code. The detector signal correction factors in multiple rod positions were calculated using Equation (10), and the relationship between the correction factor and rod position was obtained by high-order polynomial fitting. The detector signal correction factor curve is illustrated in Fig. 3. Then, the raw signals $N_{Det}(t)$ were normalized and modified using Equation

(9).

We can see from the figure that the spatial effects of the #1 and #7 control rod drops were significantly different. The detector signal correction factor of the #1 control rod changes greatly, especially when the rod drops to the bottom, while that of the #7 rod stays near 1, fluctuating within approximately 4%. Hence, the raw data of the #1 rod is considered to be "bad", and that of the #7 rod is "good".

Table 2. Dynamic reactivity correction factors

| Measured rod | $\rho_{dyn,c}$ (pcm) | $\rho_{st,c}$ (pcm) | $C_{\text{dyn}}$ |
|---|---|---|---|
| #1 rod | −4716.0 | −4874.9 | 1.0337 |
| #7 rod | −2812.3 | −2664.2 | 0.94731 |

By substituting the amplitude function $p(t)$ calculated with TAPUX into Equation (1), the calculated dynamic reactivity $\rho_{dyn,c}$ was obtained. With the static $k_{eff}$ calculated by MCNP, the static reactivity $\rho_{st,c}$ was determined. In this research, we focused on the integral rod worth; hence, the corresponding dynamic reactivity correction factor was calculated using Equation (17), and the results are listed in Table 2. The dynamic reactivity correction factors are close to 1, which means that the dynamic effect is small during the process.

3.5 Uncertainty Analysis

In references [22] and [23], the uncertainty was estimated using the standard deviation, and a first-order Taylor formula was used to linearize and calculate an approximation of the uncertainty. Based on the analytical method of uncertainty in references [22] and [23], the uncertainty of the detector signal correction factor $C_{\text{det}}$ and the dynamic reactivity correction factor $C_{\text{dyn}}$ were determined. Then, the uncertainties were propagated to calculate the final reactivity.

3.5.1 Detector signal correction factor $C_{\text{det}}$ uncertainty

According to the uncertainty definition in reference [23], the relative uncertainty was evaluated as the relative standard deviation. For the MCNP output, the relative error was evaluated as the relative standard deviation. Thus, the relative uncertainties of the detector neutron flux $\phi(r_d,t)$ and neutron flux distribution $\phi(r,E,t)$ were evaluated using the relative error in the MCNP output file.

The relative uncertainties of the physical parameters were propagated to the detector signal correction factor $C_{\det}$.

Using the analytical method of the standard deviation [22,23], the relative uncertainty of the adjoint neutron flux distribution $\phi_0^*(r,E)$ was determined.

$$\frac{u(\phi_{0,m}^*(E))}{\phi_{0,m}^*(E)} = \frac{u(I_{FP,m}^{(\lambda')})}{I_{FP,m}^{(\lambda')}} = \sqrt{\sum_{i=1}^{\lambda'} \left[\frac{u(k_{\theta_0,m}^{(i)})}{k_{\theta_0,m}^{(i)}}\right]^2}. \tag{28}$$

For each criticality calculation, we assumed that the relative uncertainties of $k_{\theta_0}^{(i)}$ from different cycles were the same because the calculation condition was unchanged even though the source points changed across cycles. Here, $u(k_{\theta_0}^{(i)})/k_{\theta_0}^{(i)}$ was assumed to be the same in one critical calculation. Then, we obtained the relative uncertainty of $I_{FP}$, [16]

$$\frac{u(I_{FP,m}^{(\lambda')})}{I_{FP,m}^{(\lambda')}} = \sqrt{\lambda'} \frac{u(k_{\theta_0,m}^{(i)})}{k_{\theta_0,m}^{(i)}}. \tag{29}$$

As the neutrons propagated, the fission neutron emission (source) produced by the progenies converged to the fundamental mode. The $k$ calculated in these fundamental mode source condition was considered to be under the same simulation conditions, and it was used to determine the relative uncertainty of $k$. [16] Therefore, we skipped 50 cycles and used the $k$ of the last 50 cycles to calculate the relative uncertainty of $\phi_0^*(r,E)$.

3.5.2 Uncertainty of dynamic reactivity correction factor $C_{\text{dyn}}$

According to the uncertainty of $k_{\text{eff}}$, which is illustrated by the MCNP results, the uncertainty of the dynamic reactivity correction factor $C_{\text{dyn}}$ was determined. Using Equation (17), the relative uncertainty of $C_{\text{dyn}}$ can be described as follows:

$$u_{rel}(C_{dyn}) = \frac{u(C_{dyn})}{C_{dyn}} = \frac{u(k_{eff})}{k_{eff}}. \tag{30}$$

Here, the uncertainty of TAPUX was ignored because it is a deterministic code.

3.5.3 Uncertainty of final reactivity modification

Based on the above calculations, the relative uncertainty of the detector signal correction factor $C_{\text{det}}$ was obtained, and the relative uncertainty curve of the detector signal correction factor was obtained with the relative uncertainties at different rod positions by fitting. Then, the relative uncertainty of the detector signal correction factor $C_{\text{det}}$ and the dynamic reactivity correction factor $C_{\text{dyn}}$ were propagated to the reactivity. Using Equations (1), (9), and (16), the relative uncertainty caused by the modification in the reactivity was determined, as illustrated in Table 3.

**4 Results and Discussion**

4.1 Results based on modified rod drop experimental technique

4.1.1 Results based on detector signal modification

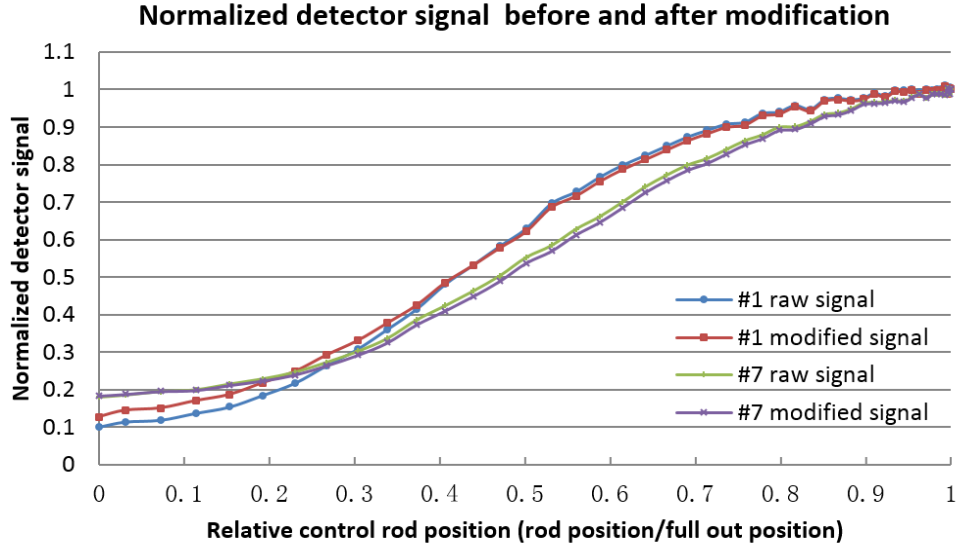

Fig. 4. Normalized detector signals before and after modification.

According to the measurement procedure described in Section 2.1, the detector signal was obtained through experimentation. Because the control rod position changes rapidly during the rod drop [24–26], the sample frequency was set to 100 Hz in the measurement. Neutron detector A measured the neutron signal of control rod cluster #1 dropping from the top to bottom, and neutron detector B measured that of control rod cluster #7. The experimental data were normalized to $N(0)$, and the data were transformed into the relationship between the signal and relative control rod position using the relationship between time and the relative control rod position during the free fall rod drop process.

The normalized detector signal $N(t)/N(0)$ is displayed in Fig. 4 (raw signal), where the relative control rod position '1' means that the rod is at the top of the core, and the relative control rod position '0' means that the rod is at the bottom of the core.

Using the signal modification procedure in Section 3, the detector signal correction factor was determined, and the modified detector signal was obtained. The normalized modified detector signal $N_c(t)/N_c(0)$ is displayed in Fig. 4 (modified signal). As illustrated in this figure, the detector signal falls rapidly during the rod drop process, which is less than 1 s, and the signal decreases by approximately one order of magnitude for the #1 rod.

By substituting the measured detector signal into the inverse kinetic equation, the reactivity was obtained. The integral rod worth calculated with both raw and modified data is listed in Table 3.

4.1.2 Results based on dynamic reactivity modification

The integral rod worth both with and without signal modification were modified with dynamic reactivity correction factor $C_{\text{dyn}}$ using Equation (16) are also displayed in Table 3.

4.1.3 Final results

With both detector signal and dynamic reactivity modifications, the final results were obtained, as exhibited in Table 3. The measurement result is compared with the MCNP result, and the relative difference with the MCNP result is listed in the table.

Table 3. Comparison of modified integral rod worth

| Results | Measured rod | #1 rod | #7 rod |
|---|---|---|---|
| MCNP code | Reactivity (pcm) | −4874.9 | −2664.2 |
| | ±Uncertainty (pcm) | ±0.2 | ±0.1 |
| Conventional rod drop experimental technique (with raw data) | Reactivity (pcm) | −5604.4 | −2803.3 |
| | Relative difference | 14.96% | 5.22% |
| Modified results 1 (with signal modification only) | Reactivity | −4288.1 | −2733.1 |
| | ±Uncertainty (pcm) | ±43.1 | ±28.1 |
| | Relative difference | −12.04% | 2.59% |
| Modified results 2 (with dynamic reactivity modification only) | Reactivity | −5793.2 | −2655.6 |
| | ±Uncertainty (pcm) | ±0.3 | ±0.1 |
| | Relative difference | 18.84% | −0.32% |
| Final reactivity (with signal and dynamic reactivity modifications) | Reactivity | −4432.5 | −2589.1 |
| | ±Uncertainty (pcm) | ±43.1 | ±28.1 |
| | Relative difference | −9.07% | −2.82% |

As the results demonstrate, the discrepancy between the conventional rod drop experimental technique and MCNP is large because of the spatial and dynamic effects in the conventional measurement, while most of the modified results are improved.

4.2 Effect of constraining the shape function

As we discussed in Section 2.2.1.2, the purpose of constraining the shape function is to reduce the reactivity perturbation caused by the calculation error of the neutron flux or shape function. The raw signal was modified using the detector neutron flux $\phi(r_d, t)$, which was calculated by MCNP without constraining the shape function, and the rest of the modifications are the same as those previously mentioned. Then, the final reactivities both with and without constraining the shape function were compared.

The investigation was based on #1 rod drop data, and the spatial and dynamic effects of the #1 rod were strong. In addition to the #1 rod drop measurement data, other #1 rod drop measurement data were considered, where the data were measured in another critical control rod pattern (critical control rod pattern 3 in Table 1). The results are compared in Table 4.

The results indicated that constraining the shape function improves the reactivity results. Although the improvement appears small and the constraint appears complex, the absolute value of the reactivity improvement is considerable for such a reactivity measurement scale. For small experimental reactors, the calculation cost of constraining the shape function is acceptable, and constraining the shape function is recommended.

Table 4. Effect of constraining shape function in #1 rod drop measurement

| Results \ Measured state | | critical pattern 1 | critical pattern 3 |
|---|---|---|---|
| MCNP code | Reactivity (pcm) | −4874.9 | −5447.8 |
| | ±Uncertainty (pcm) | ±0.2 | ±0.3 |
| Conventional rod drop experimental technique | Reactivity (pcm) | −5604.4 | −5869.5 |
| | Relative difference | 14.96% | 7.74% |
| Final reactivity (without constraining the shape function) | Reactivity (pcm) | −4373.4 | −5231.0 |
| | ±Uncertainty (pcm) | ±21.4 | ±25.2 |
| | Relative difference | −10.29% | −3.98% |
| Final reactivity | Reactivity (pcm) | −4432.5 | −5301.8 |

| (with constraining the shape function) | ±Uncertainty (pcm) | ±43.1 | ±47.0 |
| | Relative difference | −9.07% | −2.68% |

4.3 Discussion

The results of the #1 and #7 rods are substantially different. We think that this is because the critical control rod patterns and the test control rod positions are different. This causes a difference in neutron flux distribution, and thus, the reactivity worth for these rods are different. Additionally, the detector positions are significantly different, resulting in different signals and correction factors.

The results indicate that the modified rod drop experimental technique can improve both "bad" and "good" raw data, and they can act as guidance for the application of modified rod drop experimental techniques in large reactivity measurement.

The signal modification (modified results 1) demonstrated sufficient performance in improving the results. With signal modification, both the #1 and #7 rod reactivity results were closer to the MCNP calculation.

The results were only worse with dynamic reactivity modification (modified results 2). For modified results 2, the #7 rod reactivity was perfectly corrected, while the #1 reactivity worsened. We think that this is because the dynamic effect is determined by the dropped rod position and critical core configuration. The dynamic correction factor was approximately 1.03 in the #1 rod measurement. The vectors of the dynamic and spatial effects were in different directions, and that of the spatial effect was much larger than that of the dynamic effect. Thus, only the dynamic reactivity modification worsens the result. Thus, detector signal modification is important and essential, and it dominates the modifications. Only the dynamic reactivity modification was insufficient.

With both signal and dynamic reactivity modifications, the final reactivity agrees well with the MCNP result, and the relative differences are much smaller than those of the conventional rod drop experimental technique. Because the spatial and dynamic effects are small in the #7 rod drop measurement, the final reactivity does not display clear improvement compared to the results of single modification (modified results 1 and 2). The final reactivity of the #7 rod is superior to that of the conventional rod drop experimental technique.

The results obtained with both "bad" (#1 rod) and "good" raw data (#7 rod) are improved by the modified method, indicating that the modified rod drop experimental technique can reduce the spatial and dynamic measurement effects. Hence, the modified method is demonstrated to be more accurate than the conventional method and valid.

## 5 Conclusions

In this study, a modified rod drop experimental technique was proposed, and it was applied to large reactivity measurement. In the modification, static physical parameters were calculated using the MCNP code, and dynamic physical parameters were calculated using a transient code. The reactivity of the rod drop process, in which large reactivity (approximately −5500 pcm) is locally inserted, was measured using the modified rod drop experimental technique. The primary conclusions can be drawn from the results as follows:

(1) When the large reactivity is locally inserted in the rod drop, the conventional rod drop experimental technique accuracy is limited by the spatial effect and the difference between the static reactivity and dynamic reactivity. The modified rod drop experimental technique can reduce the spatial and dynamic effects in the measurement, and it is more accurate and valid.

(2) The detector signal modification is important and essential, and it dominates the modification of large reactivity measurement. Modifications based on MCNP exhibit satisfactory results.

(3) The dynamic reactivity modification is necessary for large reactivity measurement.

(4) Constraining the shape function can reduce the reactivity perturbation caused by the difference between the calculated neutron flux and its true value, and the results suggest that the modification can improve the results.

**Acknowledgments**

The authors would like to thank Guoen Fu, Dong Wei, Yongmu Yang, Shibiao Pan, Cuiyun Peng, Yonglin Zhang, Yan Guo, Jie He, Hang Zhou, Aining Deng, and Yifan Zhang of the Reactor Engineering Research Sub-institute of the Nuclear Power Institute of China for their help in the experiment related to this work.

The authors would like to thank Guoen Fu, Dong Wei, Yongmu Yang, Shibiao Pan, Cuiyun Peng, Yonglin Zhang, Yan Guo, Jie He, Hang Zhou, Aining Deng, and Yifan Zhang of the Reactor Engineering Research Sub-institute of the Nuclear Power Institute of China for their help in the experiment related to this work.